\documentclass[a4paper,12pt]{article}
\usepackage{graphicx}

\newcommand{\EQ}{\begin{equation}}
\newcommand{\EN}{\end{equation}}
\newcommand{\bea}{\begin{eqnarray}}
\newcommand{\ena}{\end{eqnarray}}
\newcommand{\bdis}{\begin{displaymath}}
\newcommand{\edis}{\end{displaymath}}
\newcommand{\vs}[1]{\vspace{#1 mm}}

\renewcommand{\a}{\alpha}
\renewcommand{\b}{\beta}

\renewcommand{\d}{\delta}

\renewcommand{\t}{\tau}
\newcommand{\ep}{\epsilon}

\newcommand{\pa}{\partial}

\newcommand{\nn}{\nonumber \\}

\begin{document}

\topmargin 0pt
\oddsidemargin - 3mm

\newcommand{\NP}[1]{Nucl.\ Phys.\ {\bf #1}}
\newcommand{\PL}[1]{Phys.\ Lett.\ {\bf #1}}
\newcommand{\CMP}[1]{Comm.\ Math.\ Phys.\ {\bf #1}}
\newcommand{\PR}[1]{Phys.\ Rev.\ {\bf #1}}
\newcommand{\PRL}[1]{Phys.\ Rev.\ Lett.\ {\bf #1}}
\newcommand{\PREP}[1]{Phys.\ Rep.\ {\bf #1}}
\newcommand{\PTP}[1]{Prog.\ Theor.\ Phys.\ {\bf #1}}
\newcommand{\PTPS}[1]{Prog.\ Theor.\ Phys.\ Suppl.\ {\bf #1}}
\newcommand{\NC}[1]{Nuovo.\ Cim.\ {\bf #1}}
\newcommand{\JPSJ}[1]{J.\ Phys.\ Soc.\ Japan\ {\bf #1}}
\newcommand{\MPL}[1]{Mod.\ Phys.\ Lett.\ {\bf #1}}
\newcommand{\IJMP}[1]{Int.\ Jour.\ Mod.\ Phys.\ {\bf #1}}
\newcommand{\AP}[1]{Ann.\ Phys.\ {\bf #1}}
\newcommand{\RMP}[1]{Rev.\ Mod.\ Phys.\ {\bf #1}}
\newcommand{\PMI}[1]{Publ.\ Math.\ IHES\ {\bf #1}}
\newcommand{\JETP}[1]{Sov.\ Phys.\ J.E.T.P.\ {\bf #1}}
\newcommand{\TOP}[1]{Topology\ {\bf #1}}
\newcommand{\AM}[1]{Ann.\ Math.\ {\bf #1}}
\newcommand{\LMP}[1]{Lett.\ Math.\ Phys.\ {\bf #1}}
\newcommand{\CRASP}[1]{C.R.\ Acad.\ Sci.\ Paris\ {\bf #1}}
\newcommand{\JDG}[1]{J.\ Diff.\ Geom.\ {\bf #1}}
\newcommand{\JSP}[1]{J.\ Stat.\ Phys.\ {\bf #1}}

\begin{titlepage}
\setcounter{page}{0}

\vs{5}
\begin{center}
{\Large  Note on a Closed String Field Theory \\
           from Bosonic IIB Matrix Model}

\vs{10}

{\large Daiji\ Ennyu\footnote{e-mail address:
ennyu@hiroshima-cmt.ac.jp}} \\
{\em 
Hiroshima National College of Maritime Technology \\ 
Toyota-Gun, Hiroshima 725-0200, Japan} \\
{\large Hiroshi\ Kawabe\footnote{e-mail address:
kawabe@yonago-k.ac.jp}} \\
{\em
Yonago National College of Technology \\ 
Yonago 683-8502, Japan} \\
and \\ 
{\large Naohito\ Nakazawa\footnote{e-mail address:
naohito@post.kek.jp}} \\
{\em 
High Energy Accelerator Research Organization(KEK) \\ 
Tsukuba, Ibaraki, 305-0801, Japan}  \\
\end{center}

\vs{8}
\centerline{{\bf{Abstract}}}

We apply stochastic quantization method to the bosonic part of IIB matrix model, i.e., a naive zero volume limit of large N Yang-Mills theory, to construct a collective field theory of Wilson loops. The Langevin equation for Wilson loops can be interpreted as the time evolution of closed string fields. The corresponding Fokker-Planck hamiltonian deduces a closed string field theory which describes interacting Wilson loops with manifest Lorentz invariance. 

\end{titlepage}
\newpage
\renewcommand{\thefootnote}{\arabic{footnote}}
\setcounter{footnote}{0}

It has been conjectured that non-perturbative definitions of superstring theories are given by supersymmetric matrix models\cite{IIA}\cite{IIBa}\cite{TypeI}. The proposal is strongly supported by the fact that the IIB matrix model\cite{IIBa} almost recovers the IIB light-cone string field theory\cite{GS}\cite{GSB} at the scaling limit\cite{IIBb}.  Although the light-cone setting is preferable to prove the equivalence, we hope to take the continuum limit of IIB matrix model preserving the manifest Lorentz invariance for studying its dynamical behaviour such as the spontaneous break down of the Lorentz invariance in the context of the superstring field theory. One way to keep the manifest Lorentz invariance is to apply stochastic quantization method (SQM)\cite{PW} to IIB matrix model. In this note, we consider only the bosonic part of IIB matrix model, which was once discussed to map the large N gauge theories to a string \cite{Bar}, to derive a hamiltonian which describes the time development of Wilson loops and to clarify how to take the continuum limit of it. The dynamical aspects of bosonic IIB matirx model have been studied\cite{HNT}\cite{AABHN} especially in relation to the $U(1)^d$ symmetry in the Eguchi-Kawai model\cite{EK}. It is pointed out that the bosonic IIB matrix model describes the weak coupling region  in the original Eguchi-Kawai model where the $U(1)^d$ symmetry is spontaneously broken\cite{BHN}, however, the eigenvalue distribution does not collapse to a point but has a finite extent at the t'Hooft limit \cite{HNT}. As a naive zero volume limit of SU(N) Yang-Mills theory, the partition function of the bosonic IIB matrix model is finite and the model appears to be well-defined in a mathematical sense \cite{AW}. Our strategy in constructing the string field theory is similar to that applied to the minimal matter matrix models\cite{JR}\cite{Na}. We mainly use SQM to construct a collective field theory of a gauge invariant observable, Wilson loops, in bosonic IIB matrix model so that we do not need any gauge fixing procedures. The Fokker-Planck hamiltonian defines a string field theoretical description of Wilson loops at the scaling limit. 

We begin with the action of the bosonic part of IIB matrix model, 
\bea
\label{eq:action}
S 
= -  {1\over4 g_0^2 N} {\rm Tr} \big([ A_\mu, A_\nu ]^2    
\big)         \ ,   
\ena
where $A_\mu$'s are hermitian $N\times N$ matrices. We define the time evolution of the fundamental matrix variables, 
$A_\mu( \t+\Delta \t ) \equiv A_\mu( \t ) + \Delta A_\mu( \t )$, 
in terms of Ito's calculus with respect to the discretized stochastic time $\t$. The time evolution, which recovers the probability distribution ${\rm e}^{-S}$ with $S$ in the equation (1) at the infinite stochastic time limit 
$\t \rightarrow \infty$, is given by the following basic Langevin equation, 
\bea
\label{eq:Langevin}
{\Delta}A_\mu(\t)
= {1\over g_0^2 N} [ A_\nu, [ A_\mu, A^\nu ]](\t) \Delta\t + \Delta\xi_\mu(\t)    \ . 
\ena
Here the white noises, $\Delta\xi_\mu$'s, are also hermitian $N\times N$ matrices. 
Their correlations are defined by 
\bea
\label{eq:noise}
<\Delta\xi_\mu(\t)_{ij} \Delta\xi_\nu(\t)_{kl}>_\xi
= 2\Delta\t \d_{\mu\nu} \d_{il} \d_{jk}   \ . 
\ena

The Wilson loop is a gauge invariant observable which is calculable in SQM without gauge fixing procedure, 
\bea
\label{eq:Wilson}
W_M ( k_1, k_2, ..., k_M ) 
=  {1\over N}{\rm Tr} \prod_{n=1}^M U_n  \ ,
\ena
where
$
U_n = {\rm exp}({i\epsilon k_n^\mu A_\mu /\sqrt{N}} )  \ . 
$
It satisfies an equation,
\bea
\label{eq:reparametrization}
{\hat k}_n^\mu \nabla^n_\mu W_M + O ( \epsilon^2 ) = 0  \ ,
\ena
where 
$
{\hat k}_n^\mu = {1\over2}( k_n^\mu + k_{n+1}^\mu )  \ ,
$
and 
$
\nabla^n_\mu = \pa /\pa ( i\epsilon k_n^\mu ) 
 -  \pa /\pa ( i\epsilon k_{n+1}^\mu )  \ .
$
As 
$
\epsilon \rightarrow 0     \ ,
$
we will consider the continuum limit such as 
$
( k_1^\mu, k_2^\mu, ..., k_M^\mu ) \rightarrow k^\mu(\sigma) \ ,
$
and 
\bea
X_n^\mu 
&=& - {\pa \over\pa ( i\epsilon k_n^\mu )} 
\rightarrow i {\delta \over \delta k^\mu(\sigma)}   \ , \nn
\nabla_n^\mu 
&=& X^\mu_{n+1} -   X^\mu_n 
\rightarrow \epsilon{X'}^\mu(\sigma)  \ ,
\ena
by introducing the world sheet coodinate $\sigma$,
$
-\pi \a \leq \sigma \leq \pi \a \ , 
$
with the \lq\lq length \rq\rq parameter of the Wilson loop,  
$
\a \equiv M\epsilon/2\pi  \ .
$
Then the equation (\ref{eq:reparametrization}) corresponds to, 
$
k^\mu(\sigma){X'}_\mu(\sigma) W_\a[k(\sigma)] = 0        \ ,
$
which ensures the reparametrization invariance of the Wilson loop with respect to the world sheet coordinate
$\sigma$\cite{IIBb}.

The generalized Langevin equation for the Wilson loop is derived from the variation 
\bea
\label{eq:variation}
\Delta W_M 
&=& {1\over N}\sum_{m=1}^M{\rm Tr}( U_1...\Delta U_m...U_M )  \nn
&+& {1\over N}\sum_{1\leq m < n \leq M} 
{\rm Tr}( U_1...\Delta U_m...\Delta U_n...U_M ) + O(\Delta\t^{3/2})  \ .
\ena
From the basic Langevin equation (\ref{eq:Langevin}) and the noise correlation (\ref{eq:noise}), we have,
\bea
\label{eq:link-Langevin}
\Delta U_n 
&=& i\int^1_0\! d\!s U_n(s) \Delta Y_n U_n (1-s)    \nn
&-& 2\epsilon^2 \Delta\t{1\over N}(k_n^\mu)^2 \int^1_0\! d\!s \int^s_0\! d\!t 
U_n(1-t){\rm Tr}U_n(t)  + O(\Delta\t^{3/2})   \ , 
\ena
where 
$
U_n (s) = {\rm exp}(isY_n)  \ ,
$
and 
$
Y_n \equiv \epsilon k_n^\mu A_\mu / \sqrt{N}  \ .
$
The first term in r.h.s. of (\ref{eq:variation}) is equivalent to the insertion of the equations of motion and the noise variables which can be evaluated by the similar procedure in Ref.\cite{IIBb}.
%
%
Then we obtain 
\bea
\label{eq:free}
{1\over N}\sum_{n=1}^M{\rm Tr}( U_1...\Delta U_n...U_M )  
&=& {1\over N}\sum_{n=1}^M{\rm Tr}( U_1...U_n\big\{ 
  i\epsilon k_n^\mu \Delta\xi_\mu /\sqrt{N}   \nn
&-& \Delta\t{3\over 2g_0^2N} (\nabla_n^\mu)^2 
- \Delta\t \epsilon^2(k_n^\mu)^2 I_n 
\big\} U_{n+1}...U_M )         \ ,
\ena
where 
$
I_n = {2\over N} \int^1_0\! d\!s \int^s_0\! d\!t 
U_n(1-t){\rm Tr}U_n(t)   \ .
$
The second term in r.h.s. of (\ref{eq:variation}) will give a splitting interaction of the Wilson loop. It reads, 
\bea
\label{eq:int}
&{}& {1\over N} \!\!\!\sum_{1\leq m < n \leq M}         
{\rm Tr}( U_1...\Delta U_m...\Delta U_n...U_M ) + O(\Delta\t^{3/2})  \nn
&=& - 2 \Delta\t \epsilon^2 {1\over N^2}\sum_{1\leq m < n \leq M} 
{\hat k}_m \cdot {\hat k}_n {\rm Tr}(U_1 ...U_mU_{n+1}...U_M)
{\rm Tr}(U_{m+1}...U_n)           \ .
\ena
We here notice that the splitting interaction terms make effects to renormalize the matrices,  $(I_n)_{ij}$'s, in equation (\ref{eq:free}) at the continuum limit. A possible contribution may come from, 
$
\sum_{1\leq n \leq M}{\rm Tr}( U_1...\Delta U_n\Delta U_{n+1}...U_M )   \ ,
$
%
%
%
and yields terms which renormalize the kinetic term $k_\mu(\sigma)^2$ at the continuum limit without changing the sign of it. In the following, we assume that 
$
(I_n)_{ij} = I \d_{ij} 
$
and the factor $I$ includes these contributions. 

Now we take the continuum limit. We introduce the renormalized (continuum) stochastic time, 
$
d\t
\equiv \ep^D \Delta\t       \ ,
$
and the renormalized Wilson loop
\bea
W_\a[k(\sigma)] \equiv 
\epsilon^{- \delta} W_M 
= {1\over N\epsilon^\delta} {\rm Tr}\prod_{n=1}^M U_n    \ . 
\ena

To take the well-defined continuum limit 
$
\epsilon \rightarrow 0  \  
$
in the generalized Langevin equation 
for Wilson loops, we require that the free part except the noise term in equation (\ref{eq:free}) is identified to the constraint of the reparametrization invariance with respect to the time coordinate $\t$ ; 
$
\Bigl( k^{\mu} (\sigma ) \Bigr)^2 
 +\left( {1 \over 2\pi \alpha ^{\prime} } X^{\prime \mu}(\sigma)\right) ^2  \ .
$
We also require for the splitting interaction part, equation (\ref{eq:int}), to be finite at the continuum limit.

From these requirements, we assume
\bea
\label{eq:double-1}
g_0^2\epsilon^{D-1} = {\a'}^2      \ , 
I \epsilon^{1-D} = 1      \ , 
\delta = 1 + D    \ ,
\ena
up to some numerical constants. Then we have,
\bea
\label{eq:continuum-Langevin}
\Delta W_{\alpha}[k(\sigma)] & = &
 \Delta_{\xi}W_{\alpha}[k(\sigma)] \nonumber\\
& - &
 d\tau \int_{-\pi \alpha}^{\pi \alpha} d\sigma
 \left\{ \Bigl( k^{\mu} (\sigma ) \Bigr)^2 
 +\left( {1 \over 2\pi \alpha ^{\prime} } X^{\prime \mu}(\sigma)\right) ^2 \right\}
 W_{\alpha }[k ( \sigma )]  \nonumber\\
& - &
 2d\tau \int_0^{\alpha} \!\!d\alpha _1 \int _{-\alpha _2}^{\alpha _2}\!\!d\beta _3 \!\int \!\!\!
 {\cal D}k^{(1)}(\sigma _1) {\cal D}k^{(2)}(\sigma _2) 
 \{\epsilon ^{1/2} \hat{k}^{(3)} (\pi \alpha _1 + \pi \beta _3 )\}
 \nonumber\\
& &
 \cdot \{\epsilon ^{1/2} \hat{k}^{(3)} (-\pi \alpha _1 + \pi \beta _3 )\}
 \delta (1,2,3_{\beta _3}) W_{\alpha _1} [k^{(1)}(\sigma _1)]
 W_{\alpha _2} [k^{(2)}(\sigma _2)]
\ena
where $\a = \a_1 + \a_2$. The overlap condition for the splitting interaction is the following, 
\begin{center}
\begin{tabular}{lll}
$k^{(3)}(\sigma _3) - \theta _1 k^{(1)} (\sigma _1)
 -\theta _2 k^{(2)} (\sigma _2) =0 $
& \hspace*{2em} &
$\theta _1 = \theta (\pi \alpha _1 - |\sigma |) $\\
&&
$\theta _2 = \theta (|\sigma | - \pi \alpha _1 ) $\\
&&
$\theta _3 = \theta _1 + \theta _2 = 1 $
\end{tabular}
\end{center}
\begin{center}
\begin{tabular}{lll}
$\sigma _1 = \sigma$ & \hspace*{1em} & ~$-\pi \alpha _1 \leq \sigma
 \leq \pi \alpha _1 $\\
$\sigma _2 = \left\{
  \begin{array}{l}
  \sigma -\pi \alpha _1 \\
  \sigma +\pi \alpha _1
  \end{array} \right.$
 & &
$ \begin{array}{l}
  \pi \alpha _1 \leq \sigma \leq \pi (\alpha _1 + \alpha _2)\\
  -\pi (\alpha _1 + \alpha _2 ) \leq \sigma \leq -\pi \alpha _1
  \end{array} $\\
$\sigma _3 = \sigma + \pi \beta_3$
 & &
~$ -\pi ( \alpha _1 + \alpha _2) \leq \sigma \leq \pi (\alpha _1 + \alpha _2) $
\end{tabular}
\end{center}
\begin{equation}
\label{eq:split-overlap}
-\alpha _2 \leq \beta _3 \leq \alpha _2
\end{equation}
The parameter $\b_3$ is introduced to represent the splitting point which is specified by 
$
\sigma_3 = \pm \pi\a_1 + \pi\b_3   
$
on the Wilson loop. We denote the overlap condition as 
$
\delta (1,2,3_{\b_3})   
$.
%
%
%
The derivative coupling at the splitting interaction point is defined by 
$
{\hat k}_\mu^{(3)}(\pm \pi\a_1 + \pi\b_3)  
= {1\over2}\big\{ k_\mu^{(3)}(\pm \pi\a_1 + \pi\b_3 + \epsilon ) 
+ k_\mu^{(3)}(\pm \pi\a_1 + \pi\b_3 - \epsilon )\big\}  \   .
$
Since this quantity diverges as $\epsilon^{-1/2}$ at the interaction point\cite{GSB}, the factor $\epsilon^{+1/2}$ is necessary in the equation (\ref{eq:continuum-Langevin}) for the well-defined continuum limit. 

The noise variable is also defined by,
\bea
\label{eq:continuum-noise}
\Delta _{\xi}W_{\alpha }[k(\sigma)] = {1 \over N \epsilon ^{\delta}} \sum_{n=1}^M
 {\rm Tr} \left( U_1 \cdot \cdot \cdot U_n
\left\{ i\epsilon \hat{k}_n^{\mu}
 \Delta \xi _{\mu} /\sqrt{N} \right\} U_{n+1} \cdot \cdot \cdot U_M \right)
\ena
We obtain the following correlation of them at the continuum limit.
\bea
\label{eq:continuum-noise-co}
&& <\Delta _{\xi} W_{\alpha _1} [k^{(1)}(\sigma _1)]
 \Delta _{\xi} W_{\alpha _2} [k^{(2)}(\sigma _2)]  > \nonumber\\
&& \hspace*{3em}
= -2d \tau G_{\rm st} \int_{-\alpha _1}^{\alpha _1}\pi d\beta _1
 \int _{-\alpha _2} ^{\alpha _2}\pi d\beta _2
 \int\!\!\!{\cal D}k^{(3)}(\sigma _3) \nonumber\\
&&  \hspace*{3em}
 \times
 \{\epsilon ^{1/2} \hat{k}^{(1)}(\pi \alpha _1 + \pi \beta _1)\}
  \{\epsilon ^{1/2} \hat{k}^{(2)}(\pi \beta _2)\}
  \delta (1_{\beta _1},2_{\beta _2},3)
 W_{\alpha _3}[k^{(3)}(\sigma _3)]
\ena
where $\a_3 = \a_1 + \a_2$. Here we have identified 
\bea
\label{eq:double-2}
N^2\epsilon^{2+2D} = 1/G_{st}       \ . 
\ena
We notice that, for the merging interaction, the overlap condition is slightly different from that of splitting interaction. 
$$k^{(3)}(\sigma _3) - \theta _1 k^{(1)} (\sigma _1)
 -\theta _2 k^{(2)} (\sigma _2) =0 $$
\begin{center}
\begin{tabular}{lll}
$\sigma _1 = \sigma +\pi \beta _1$ & \hspace*{1em} & ~$-\pi \alpha _1 \leq \sigma
 \leq \pi \alpha _1 $\\
$\sigma _2 = \left\{
  \begin{array}{l}
  \sigma + \pi \beta _2 -\pi \alpha _1 \\
  \sigma + \pi \beta _2 +\pi \alpha _1
  \end{array} \right.$
 & &
$ \begin{array}{l}
  \pi \alpha _1 \leq \sigma \leq \pi (\alpha _1 + \alpha _2)\\
  -\pi (\alpha _1 + \alpha _2 ) \leq \sigma \leq -\pi \alpha _1
  \end{array} $\\
$\sigma _3 = \sigma $
 & &
~$ -\pi ( \alpha _1 + \alpha _2) \leq \sigma \leq \pi (\alpha _1 + \alpha _2)$
\end{tabular}
\end{center}
\begin{equation}
\label{eq:merge-overlap}
-\alpha _1 \leq \beta_1 \leq \alpha _1,~~~-\alpha _2 \leq \beta_2 \leq \alpha _2
\end{equation}
The overlap $\delta$-function is now denoted by 
$
\delta (1_{\b_1},2_{\b_2},3)  \ , 
$
The merging interaction point is specified by 
$
\sigma_1 = \pm \pi\a_1 + \pi\b_1 
$
and
$
\sigma_2 = \pi\b_2 \ .
$
The derivative coupling at the merging interaction point is also different from that at the splitting interaction point. They are given by
\bea
{\hat k}^{(1)}( \pi\a_1 + \pi\b_1)  
&=& {1\over2}( {\hat k}^{(3)}(\pi\a_1 - \epsilon ) 
+ {\hat k}^{(3)}( - \pi\a_1 + \epsilon ))      \ , \nn
{\hat k}^{(2)}( \pi\b_2)  
&=& {1\over2}( {\hat k}^{(3)}( \pi\a_1 + \epsilon ) 
+ {\hat k}^{(3)}( - \pi\a_1 - \epsilon ))       \ .
\ena
\begin{figure}[htbp]
\begin{minipage}{.45\linewidth}
	\caption{splitting interaction}
	\includegraphics[width=\linewidth]{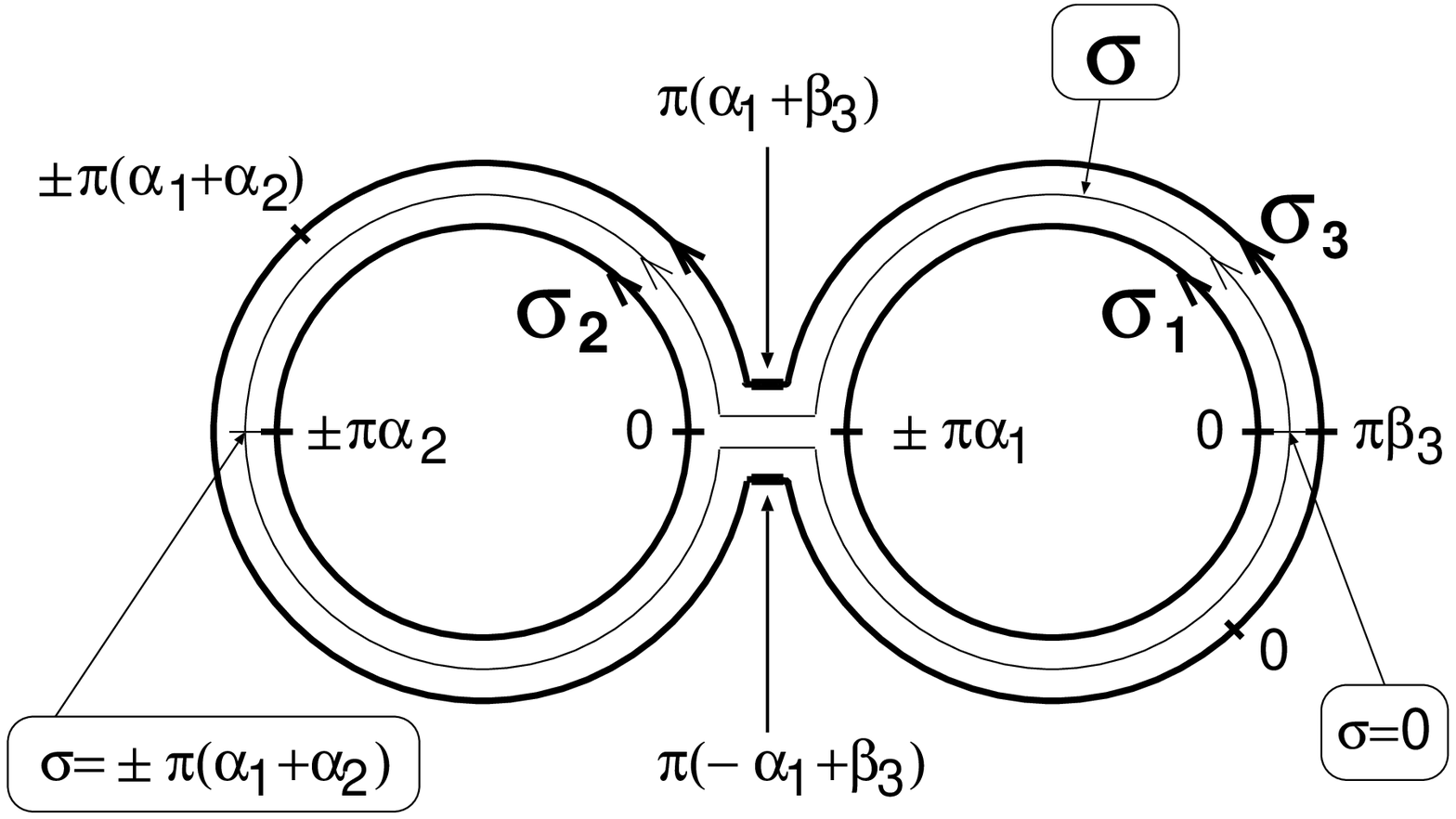}
	\label{fig:splitting}
\end{minipage}
\begin{minipage}{.45\linewidth}
	\caption{merging interaction}
	\includegraphics[width=\linewidth]{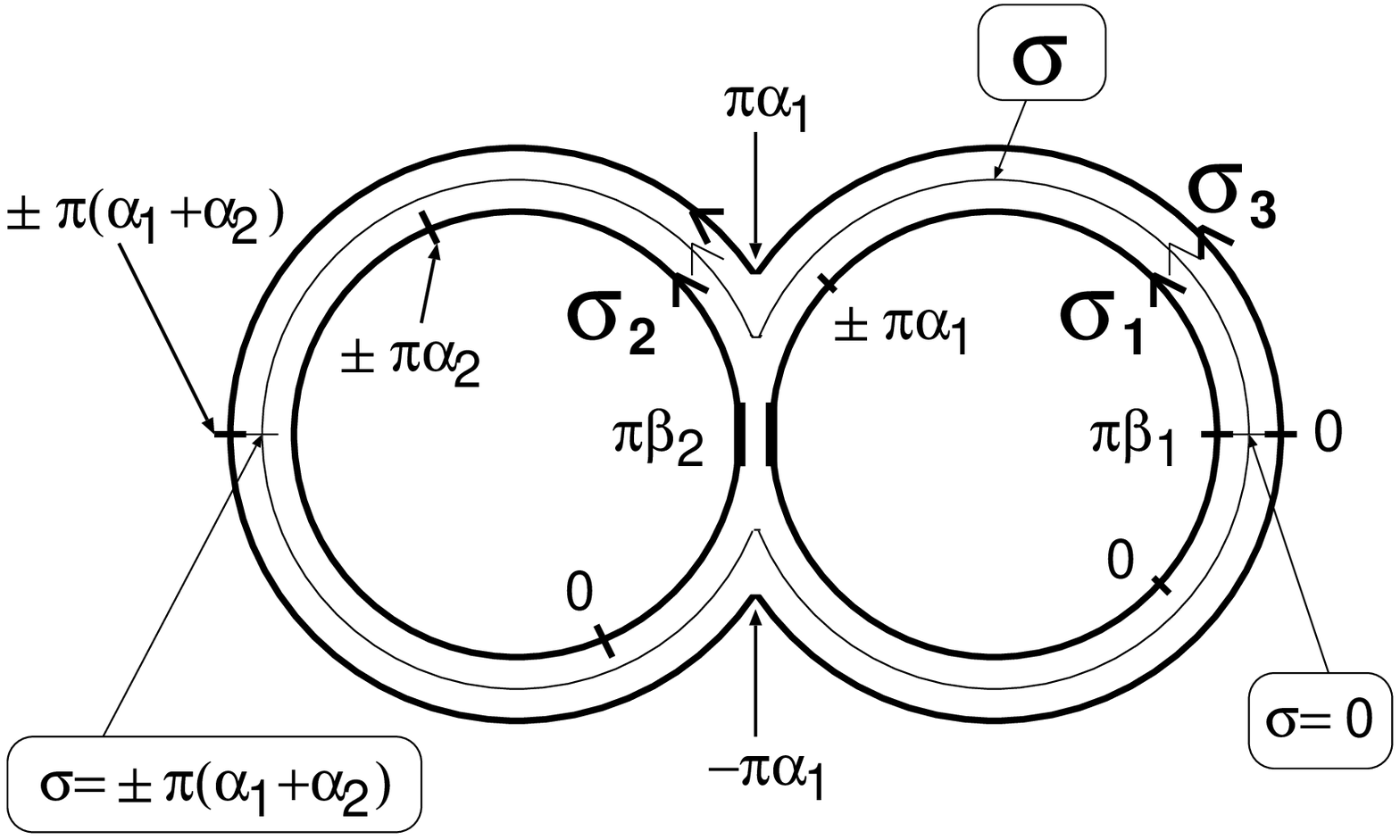}
	\label{fig:merging}
\end{minipage}
\end{figure}

Before deriving the Fokker-Planck hamiltonian, we add some remarks on the scaling limit which is defined by equations (\ref{eq:double-1}) and (\ref{eq:double-2}). Since we fix the string tension $\a'$ and the string coupling constant $G_{st}$ to be finite at the large N limit, the equations (\ref{eq:double-1}) and (\ref{eq:double-2}) reads, 
\bea
\epsilon &\approx &  N^{- 1/(1+D)}      \ , \nn
g_0^2 &\approx & N^{- (1-D)/(1+D)}   \ .
\ena
A well-defined continuum limit requires 
$
\epsilon \rightarrow 0  \ 
$
at 
$
N \rightarrow \infty  \
$
which ensures the scaling dimension of the stochastic time $\t$ to satisfy 
$
1 + D  > 0   \ .
$
We here require a stronger condition 
$
D > 0  \
$
to obtain the smooth continuum limit for the time development of the stochastic process. By redefining the fundamental matrix variable $A_\mu$ to $A_\mu \sqrt{N}$ both in the action (\ref{eq:action}) and the definition of the Wilson loop (\ref{eq:Wilson}), the coupling $g_0^2$ in (\ref{eq:action}) is replaced by $g_0^2/N$ which is equivalent to the original coupling constant of IIB matrix model $g^2$. According to the correspondence $g_0^2 = g^2 N$, the space-time extent is expected to be characterized by the quantity 
$
R \approx \sqrt{g} N^{1/4} = \sqrt{g_0}  \  
$
\cite{HNT}. It seems that, if a continuum limit may exist, there are three possible cases, 
$ 0 < D < 1 $, $ D = 1 $ and $ D > 1 $, 
which distinguish the extent of eigen value distribution, i.e., the space-time extent. For $ 0 < D < 1 $, since 
$
R \rightarrow 0  \
$
at 
$
N \rightarrow \infty  \ ,
$
 the target space-time is collapsed. The case $D=1$ in (\ref{eq:double-1}) realizes the t'Hooft limit and it deduces $N\epsilon^2 = 1/G_{st}$. The condition was discussed in the light-cone setting\cite{IIBb}. In this case, the space-time may have a finite extent even at the large N limit. Then we may consider a dual picture to interpret the momentum $k^\mu (\sigma)$ for the closed string by assuming the boundary condition for the target space time coordinte \cite{AIKKTT}, 
$
n^\mu R = x^\mu (2\pi) - x^\mu (0) \ ,
$
with $n^\mu$, the winding number in the $\mu$ direction. The last possibility is to consider the limit $R \rightarrow \infty$ which means $D>1$. In any cases, a remarkable fact is that since 
$
N\epsilon^\delta = 1/\sqrt{G_{st}}   \ 
$
is fixed, the intrinsic Wilson loop, 
$
{\rm Tr}\prod_{n=1}^M U_n   \ ,
$
is not scaled at the scaling limit. Our analysis is restricted to the bosonic part of IIB matirx model, however, the scaling relations, (\ref{eq:double-1}) and (\ref{eq:double-2}), hold even for full IIB matirx model. The dynamical behaviour of the bosonic IIB matrix model is expected to be quite different from that of full IIB matrix model and it should be determined from the large N behaviour of the numerical factor 
$I \approx \epsilon^{D-1} $. If we assume that the factor $I$ is a function of 
$\epsilon, g_0^2, N$; $I=I(\epsilon, g_0^2, N)$, then the condition to determine the index $D$ is given by 
$
g_0^2 I(N^{-1/(1+D)}, g_0^2, N) \approx O(1)  \ .
$
Further investigation of the dynamical behaviour of this numerical factor, however, is beyond the scope of this note. 

In our approach, the stochastic process of the Wilson loop, which is described by the generalized Langevin equation, is interpreted as the time evolution in a string field theory. The corresponding Fokker-Planck hamiltonian defines a bosonic closed string field theory. In terms of the expectation value of an observable $O( W_\a )$, a function of $W_\a$'s, the Fokker-Planck hamiltonian operator ${\hat H}_{FP}$ is given by\cite{Na}
\EQ
\label{eq:def-H}
<{\bar W}_\a (0)| {\rm e}^{- \t {\hat H}_{FP} } O({\hat W_\a})|0>
\equiv <O( {W_\a}(\t)_{\xi})>_{\xi}                  \  .
\EN
In r.h.s., ${W_\a}(\t)_{\xi}$ denotes the solutions of the Langevin equation with the appropriate initial configuration ${{\bar W}_\a} (0)$. In l.h.s.,  ${\hat H}_{FP}$ is given by the differential operator in the well-known Fokker-Planck equation for the expectation value of the observable $O({W_\a})$ by replacing the closed string variable $W_\a[k(\sigma)]$ to the creation operator ${\hat W}_\a[k(\sigma)]$ and the differential ${\delta /\delta W_\a[k(\sigma)]}$ to the annihilation operator ${\hat \Pi}_\a[k(\sigma)]$. 
The commutation relations are 
\bea
\big[ {\hat \Pi}_\a [k(\sigma )] ,  {\hat W}_{\a'}[k'(\sigma' )]  \big]
= \d (\a - \a') \prod_{-\pi\a \leq \sigma \leq \pi\a}
\d ( k(\sigma) - k'(\sigma) )              \  .
\ena
At the scaling limit, we obtain the continuum Fokker-Planck hamiltonian, ${\hat H}_{FP}$, 
\bea
\label{eq:continuum-H}
\hat{H}_{\rm FP} & \!\!\!= & \!\!\!
 \int_0^{\infty} d\alpha \int\!\!\!{\cal D}k(\sigma)
 \int_{-\pi \alpha}^{\pi \alpha} d\sigma
 \left\{ \Bigl( k^{\mu} (\sigma ) \Bigr)^2 
 +\left( {1 \over 2\pi \alpha ^{\prime} } X ^{\prime \mu}(\sigma)\right) ^2 \right\}
 W_{\alpha }[k ( \sigma )] \Pi _{\alpha} [k(\sigma)] \nonumber\\
& \!\!\!+ & \!\!\!
 2 \int _0^{\infty} d\alpha
 \int_0^{\alpha}d\alpha _1 \int _{-\alpha _2}^{\alpha _2}d\beta _3
 \int\!\!\!{\cal D}k^{(1)}(\sigma _1) {\cal D}k^{(2)}(\sigma _2) {\cal D}k^{(3)}(\sigma _3)
 \{\epsilon ^{1/2} \hat{k}^{(3)} (\pi \alpha _1 + \pi \beta _3 )\} \nonumber\\
& & \hspace*{2em}
 \cdot \{\epsilon ^{1/2} \hat{k}^{(3)} (-\pi \alpha _1 + \pi \beta _3 )\}
 \delta (1,2,3_{\beta _3}) W_{\alpha _1} [k^{(1)}(\sigma _1)]
 W_{\alpha _2} [k^{(2)}(\sigma _2)] \Pi _{\alpha _3}[k^{(3)}(\sigma _3)]
\nonumber\\
& \!\!\!+ & \!\!\!
 G_{\rm st} \int _0^{\infty} \!\!\!\!d\alpha _1
 \int _0^{\infty} \!\!\!\!d\alpha _2
 \int_{-\alpha _1}^{\alpha _1} \!\!\!\!d\beta _1 
 \int _{-\alpha _2} ^{\alpha _2} \!\!\!\! d\beta _2
 \int\!\!\!{\cal D}k^{(1)}(\sigma _1) {\cal D}k^{(2)}(\sigma _2) {\cal D}k^{(3)}(\sigma _3)
  \{\epsilon ^{1/2} \hat{k}^{(1)}(\pi \alpha _1 + \pi \beta _1)\}
 \nonumber\\ 
& & \hspace*{2em} 
 \cdot  \{\epsilon ^{1/2} \hat{k}^{(2)}(\pi \beta _2)\}
 \delta (1_{\beta _1},2_{\beta _2},3)
 W_{\alpha _3}[k^{(3)}(\sigma _3)] \Pi _{\alpha _1}[k^{(1)}(\sigma _1)]
 \Pi _{\alpha _2}[k^{(2)}(\sigma _2)] 
\ena
By redefining the momentum integration variable, 
$
{\tilde k}^{(3)}({\tilde \sigma}_3) 
\equiv k^{(3)}(\sigma_3 ) = k^{(3)}(\sigma+ \pi\b_3 ) 
$
with  
$
{\tilde \sigma}_3 \equiv \sigma, (-\pi(\a_1+\a_2) \leq \sigma \leq \pi(\a_1+\a_2) ) \ , 
$
we can eliminate the $\b_3$ dependence in the splitting overlap, 
$
\delta (1,2,3_{\b_3})  \ ,
$
and in the derivative coupling, 
$
\hat{k}^{(3)} (\pm \pi \alpha _1 + \pi \beta _3 ) 
$
. We can also eliminate the $\b_1$ and $\b_2$ dependence in the merging interaction. After the integration with respect to the angle variables $\b_i, (i=1,2,3)$, the Fokker-Planck hamiltonian finally takes the form, 
\bea
\label{eq:continuum-H2}
\hat{H}_{\rm FP} &
 = & \int_0^{\infty} d\alpha \int\!\!\!{\cal D}k(\sigma)
\int_{-\pi \alpha}^{\pi \alpha} d\sigma
 \left\{ \Bigl( k^{\mu} (\sigma ) \Bigr)^2 
 +\left( {1 \over 2\pi \alpha ^{\prime} } X ^{\prime \mu}(\sigma)\right) ^2 \right\}
 W_{\alpha }[k ( \sigma )] \Pi _{\alpha} [k(\sigma)] \nonumber\\
& + & 
2 \int _0^{\infty} d\alpha
 \int_0^{\alpha}d\alpha _1 (2\pi \alpha _2)
 \int\!\!\!{\cal D}k^{(1)}(\sigma _1) {\cal D}k^{(2)}(\sigma _2) {\cal D}k^{(3)}(\sigma _3)
 \{\epsilon ^{1/2} \hat{k}^{(3)} (\pi \alpha _1)\}  \nonumber\\ 
& & \hspace*{3em}
 \cdot
 \{\epsilon ^{1/2} \hat{k}^{(3)} (-\pi \alpha _1)\}
 \delta (1,2,3) W_{\alpha _1} [k^{(1)}(\sigma _1)]
 W_{\alpha _2} [k^{(2)}(\sigma _2)] \Pi _{\alpha _3}[k^{(3)}(\sigma _3)]
\nonumber\\
& + & 
 G_{\rm st} \int _0^{\infty} d\alpha _1
 \int _0^{\infty} d\alpha _2
 (4\pi ^2 \alpha _1 \alpha _2)
 \int\!\!\!{\cal D}k^{(1)}(\sigma _1) {\cal D}k^{(2)}(\sigma _2) {\cal D}k^{(3)}(\sigma _3)
  \{\epsilon ^{1/2} \hat{k}^{(1)}(\pi \alpha _1)\}
 \nonumber\\ 
& & \hspace*{3em} 
 \cdot  \{\epsilon ^{1/2} \hat{k}^{(2)}(0)\}
 \delta (1,2,3)
 W_{\alpha _3}[k^{(3)}(\sigma _3)] \Pi _{\alpha _1}[k^{(1)}(\sigma _1)]
 \Pi _{\alpha _2}[k^{(2)}(\sigma _2)] 
\ena
where the overlap $\delta$-function $\delta (1,2,3)$ is given by the equation 
(\ref{eq:split-overlap}) with $\b_3 = 0$. Remarkably, the hamltonian consists of the splitting and the merging interaction terms where only three strings interact. At the weak string coupling limit $G_{st} \rightarrow 0$, which is equivalent to the large N limit with finite $\epsilon$, the merging interaction vanishes. 

In the definition of the string field theory interpretation (\ref{eq:def-H}), the initial state is defined by  
$
<{\bar W}_\a (0)| = <0| {\rm exp}\big\{ \int_0^\infty \!\!d\a 
 {\cal D}k\big( \sigma){\bar W}_\a[k(\sigma)]{\hat \Pi}_\a[k(\sigma)] \big\} 
  \ ,
$
where the initial configuration must be the solution of the large N factorized Schwinger-Dyson equation ( or the loop equation )\cite{Na}. The S-D equation is given by the limit $\Delta W_\a \rightarrow 0$ in the generalized Langevin equation (\ref{eq:continuum-Langevin}),
\bea
\label{eq:continuum-SD} 
& &\int_{-\pi \alpha}^{\pi \alpha} \!\!d\sigma
 \left\{ \Bigl( k^{\mu} (\sigma ) \Bigr)^2    
 + \left( {1 \over 2\pi \alpha ^{\prime} } 
 X^{\prime \mu}(\sigma)\right) ^2 \right\}
 <W_{\alpha }[k ( \sigma )]>             \nonumber\\
& + &
 2\int_0^{\alpha} \!\!d\alpha _1 \int _{-\alpha _2}^{\alpha _2}\!\!d\beta _3
 \int\!\!\!{\cal D}k^{(1)}(\sigma _1) {\cal D}k^{(2)}(\sigma _2) 
 \{\epsilon ^{1/2} \hat{k}^{(3)} (\pi \alpha _1 + \pi \beta _3 )\}
                          \nonumber\\
& &
 \cdot \{\epsilon ^{1/2} \hat{k}^{(3)} (-\pi \alpha _1 + \pi \beta _3 )\}
 \delta (1,2,3_{\beta _3}) <W_{\alpha _1} [k^{(1)}(\sigma _1)]
 W_{\alpha _2} [k^{(2)}(\sigma _2)]>   = 0  \ .
\ena
It is clear from the correlation in (\ref{eq:continuum-noise-co}) that the leading finite N effect on the correlation functions is given by the order of $1/N^2$\cite{HNT} and the S-D equation is factorizable at large N. 

In conclusion, we have derived the continuum Fokker-Planck hamiltonian from the stochastic process defined by the bosonic part of IIB matrix model. The structure of the splitting and the merging interactions we have derived are covariantized versions of those in the bosonic part of the type IIB Green-Schwarz superstring field theory in the light-cone gauge. The Fokker-Planck hamiltonian may be regarded as a closed string field theory hamiltonian which describes the time development of Wilson loops. We have constructed the field theory hamiltonian such that it does not have a free theory and the splitting interaction remains in the weak string coupling limit 
$
G_{st} \rightarrow 0    \ .
$
 The situation is analogous to the string field theories constructed from the old-fashioned matrix models. 
 
Now we comment on some issues which remains to be studied. If we assume that the bosonic IIB matrix model describes the weak coupling region where $U(1)^d$ symmetry is broken spontaneously, the Schwinger-Dyson equations in this model do not recover the planar perturbation theory defined by the continuum Yang-Mills theory\cite{GK}.  We also do not have evidence that implies the equivalence of the Fokker-Planck hamiltonian to collective field theories of Wilson loops in the canonical formalism of standard lattice gauge theories\cite{Sakita} or to the bosonic closed string field theory\cite{KK}. In a formal sense, the Fokker-Planck hamiltonian is a manifestly Lorentz invariant hamiltonian constraint for observables such as correlation functions of Wilson loops at the infinite stochastic time $\tau \rightarrow \infty$. 
\EQ
\label{eq:H-Pconstraint}
< H_{FP} \Big\{ W_{\a_1}[k^{(1)}(\sigma)], W_{\a_2}[k^{(2)}(\sigma)], ..., 
W_{\a_n}[k^{(n)}(\sigma)]) \Big\} > = 0   \  .
\EN
Thus, in light-cone variables, the constraint may be reduced to a light-cone hamiltonian as expected from the fact that the IIB matrix model almost recovers the light-cone IIB superstring field theory\cite{IIBb}. Another possibility is, 
by reversing the operator ordering of the Fokker-Planck hamiltonian, we obtain the Lorentz invariant Fokker-Planck equation for the probability distribution functional of the Wilson loop $P( W_\a [k(\sigma)] )$. 
\EQ
\label{eq:H-Pconstraint}
{\pa \over\pa \t}P( W_\a [k(\sigma)] ) = - {\tilde H}_{FP}( W_\a [k(\sigma)], 
{\frac{\d}{\d W_\a [k(\sigma)] }} )P( W_\a [k(\sigma)] )  \  .
\EN
The solution of this constraint, i.e., the Boltzmann weight of the path-integral representation defined at the infinite stochastic time, may specify a classical action of interacting Wilson loops. 

The question whether the present continuum limit is realized dynamically or not is yet to be studied, the behavior of the bosonic IIB matirx model may be clarified by evaluating the small string contribution to the kinetic term $(k_\mu(\sigma))^2$. For IIB matrix model, we expect the situation will be changed because of the existence of the stable perturbative vacuum and the weak string coupling limit of it may remain a free theory.  We hope to report on the analysis of full IIB matrix model in future publication. \\

\noindent
{\bf Acknowledgements}

The work (N.N.) is supported in part by the Ministry of Education, Science and Culture of Japan, Grant-in-Aid for Scientific Research (B), No.13135216. 
N.N. would like to thank Y. Kitazawa and S. Iso for discussions and comments, and all members in theory group at KEK for their hospitality.


\end{document}